\documentclass[preprint,showpacs,prd,endfloats*,nofootinbib]{revtex4-1}
\usepackage{graphicx}
\usepackage{amssymb}
\usepackage{amsmath}
\usepackage{bm}
\usepackage{natbib}
\newcommand{\p}[1]{(\ref{#1})}
\def\Journal#1#2#3#4{{#1} {\bf #2}, #3 (#4)}

\def\npa{{Nucl. Phys.} A}

\def\plb{{Phys. Lett.}  B}
\def\prl{Phys. Rev. Lett.}
\def\prc{{Phys. Rev.} C}
\def\prd{{Phys. Rev.} D}
\def\aap{Astron. Astrophys.}
\def\apj{Astrophys. J.}

\def\jpg{{J. Phys.} G}

\def\pr{Phys. Rep.}

\def\nat{Nature}
\def\ijmpa{Int. J. Mod. Phys. A}

\def\epjc{Eur. Phys. J. C}
\def\cqg{Classical Quantum Gravity}
\def\epl{Europhys. Lett.}
\def\ptps{Prog. Theor. Phys. Supplement}
\def\jmp{J. Math. Phys.}
\def\appb{Acta. Phys. Pol. B}
\def\pram{Pramana}
\def\grg{Gen. Rel. Grav.}
\def\mnras{Mon. Not. R. Astron. Soc.}
\def\jhep{JHEP}
\def\lnp{Lect. Notes in Phys.}
\def\sci{Science}
\def\jpcs{J. Phys. Conf. Ser.}
\def\past{Prob. Atomic Sci. Technol.}
\begin{document}
\title[]{General relativistic polytropes in anisotropic stars
}
\author{A. A.  \surname{Isayev}}
\email{isayev@kipt.kharkov.ua}
 \affiliation{Kharkov
Institute of Physics and Technology, Academicheskaya Street 1,
 Kharkov, 61108, Ukraine }
%

\date{\today}

\begin{abstract}
Spherically symmetric relativistic stars with the polytropic
equation of state (EoS), which possess the local pressure
anisotropy, are considered within the framework of general
relativity. The generalized Lane-Emden equations are derived for the
arbitrary anisotropy parameter $\Delta=p_t-p_r$ ($p_t$ and $p_r$
being the transverse and radial pressure, respectively). They are
then applied to some special ansatz for the anisotropy parameter
 in the form of the differential relation between the
anisotropy parameter $\Delta$ and the metric function $\nu$. The
analytical solutions of the obtained equations are found for
incompressible fluid stars and then used for getting their
mass-radius relation, gravitational and binding energy. Also,
following the Chandrasekhar variational approach, the dynamical
stability of  incompressible anisotropic fluid stars with the
polytropic EoS against radial oscillations is studied. It is shown
that the local pressure anisotropy with $p_t>p_r$ can make the
incompressible fluid stars unstable with respect to radial
oscillations, in contrast to incompressible isotropic fluid stars
with the polytropic EoS which are dynamically stable.
\end{abstract}



\maketitle

\section{Introduction}
\label{I}

It was argued in Ref.~\cite{ApJ74Bowers} that, despite the
spherically symmetric distribution of matter inside a compact
stellar object, it can be characterized by the local pressure
anisotropy. The analysis of the generalized equations of hydrostatic
equilibrium, allowing for the pressure anisotropy, shows that
anisotropy may have the substantial effect on the maximum
equilibrium mass and gravitational surface
redshift~\cite{ApJ74Bowers,CQG11Horvat,PRD11Harko}. At not too high
densities, the impact of anisotropy can be studied within Newtonian
gravity~\cite{PRD13'87Herrera,EPJC15Shojai}. At higher densities
($\varrho \gtrsim 10^{15}$~g/cm$^3$), both the relativistic effects
and the effects of general relativity become
important~\cite{ARAA72Ruderman,PR97Herrera,PRD04Herrera,CQG05Cattoen,PRD13Herrera,EPJC16Maurya,EPJC16Shojai}.
In order to study the influence of the local pressure anisotropy on
the specific basis, it is necessary to know the concrete physical
reasons responsible for its appearance, such as, e. g., the
existence of a solid core~\cite{ARAA72Ruderman,PR97Herrera},
occurrence of spontaneous deformation of Fermi
surfaces~\cite{PRL02Muther,PRC03Muther}, availability of superfluid
states with the finite orbital momentum of Cooper
pairs~\cite{PTP93Takatsuka,PRC98Baldo,PRC02IR,NPA03Zverev,EPL08Zuo}
or finite superfluid momentum~\cite{PRC02I,RMP04Casalbuoni}, or
 the presence of strong
magnetic fields inside a
star~\cite{Kh,FIKPS,PRD11Paulucci,LNP13Ferrer,IY_PRC11,IY_PLB12,IY_PAST12,JPG13IY,IJMPA14I,PRC15I,JPCS15I,PRD14Chu,PRD15Chu}.

With account of the pressure anisotropy, the equation of state (EoS)
of the system will be also necessarily anisotropic. The EoS is the
essential ingredient in solving the equations of the hydrostatic
equilibrium and its importance was underlined in
Ref.~\cite{PRD10Varela}. In the given work, we choose the polytropic
EoS, which is widely used in many astrophysical
applications~\cite{ApJ64Tooper,ApJ65Tooper,Horedt04,JMP98Nilsson,CQG03Heinzle,APPB07Kinasiewicz,PRD09Read}.
For the polytropic index $n=2$, the exact solutions of Einstein
equations with the pressure anisotropy were obtained in
Ref.~\cite{Pram12Thirukkanesh}. Anisotropic spheres with the uniform
energy density in general relativity were studied in
Refs.~\cite{GRG89Maharaj,EPJC16Shojai}, and with the variable energy
density in Ref.~\cite{GRG94Gokhroo}. Note also that the exact
solutions of Einstein equations for spherical anisotropic stars with
the linear EoS were obtained in Ref.~\cite{MNRAS07Sharma} for some
particular mass distribution.

In the present work, we will study spherically symmetric
relativistic anisotropic stars with the polytropic EoS, aiming to
obtain the generalized Lane-Emden equations for the special ansatz
for the anisotropy parameter in the form of the differential
relation between the anisotropy parameter and the metric function
$\nu$. This approach is different from that suggested in
Refs.~\cite{ApJ74Bowers,Pram12Thirukkanesh} which consists in
setting the anisotropy parameter, or some metric function, in the
specific functional form. In general case, the obtained Lane-Emden
equations can be integrated only numerically, but the analytic
solutions can be found for incompressible fluid stars which  are
then used to get their mass-radius relation, gravitational and
binding energy. Also, we apply the Chandrasekhar variational
procedure~\cite{ApJ64Chandrasekhar} to study the dynamical stability
of  incompressible anisotropic fluid stars with respect to radial
oscillations.

\section{Basic equations}
For spherically symmetric stars, the line element is written in the
form
\begin{align}
ds^2=e^{\nu(r,t)} dt^2-e^{\lambda(r,t)}
dr^2-r^2(d\theta^2+\sin^2\theta d\varphi^2),\label{le}
\end{align}
where we use the system of units with $c=1$. While the matter
distribution inside a star is spherically symmetric, we allow the
existence of the local pressure anisotropy in its interior with the
different values of the radial $p_r$ and transverse
$p_\theta=p_\varphi\equiv p_t$ pressures. The anisotropic
energy-momentum tensor for the static configuration reads
\begin{align}
T_i^k=\mathrm{diag}(\varepsilon,-p_r,-p_t,-p_t),
\label{emt}\end{align} where $\varepsilon$ is the energy density of
the system. The space-time geometry and matter distribution are
related by  Einstein equations:
\begin{align}
R_i^k-\frac{1}{2}R\delta_i^k=8\pi GT_i^k.\label{Eeq}
\end{align}
With account of Eqs.~\p{le}, \p{emt}, Einstein equations read
\begin{align}
e^{-\lambda}\biggl(\frac{\lambda\,'}{r}-\frac{1}{r^2}\biggr)+\frac{1}{r^2}=8\pi
G\varepsilon,\label{ve}\\
e^{-\lambda}\biggl(\frac{\nu\,'}{r}+\frac{1}{r^2}\biggr)-\frac{1}{r^2}=8\pi
Gp_r,\label{pr}\\
\frac{1}{2}e^{-\lambda}\biggl(\nu\,{''}-\frac{1}{2}\nu\,'\lambda\,'+\frac{1}{2}\nu\,'^2+
\frac{\nu\,'-\lambda\,'}{r}\biggr)=8\pi Gp_t, \label{pt}\end{align}
where prime denotes the differentiation with respect to~$r$. From
Eqs.~\p{ve}--\p{pt}, or, equivalently, from the vanishing divergence
of the energy-momentum tensor $T_{i;k}^k=0$, one can get the
equation for the hydrostatic equilibrium in the presence of the
pressure anisotropy in the form
\begin{align}
p_r^{\,\prime}=-\frac{\nu\,'}{2}(\varepsilon+p_r)+\frac{2\Delta}{r},
\quad \Delta\equiv p_t-p_r, \label{prpr}
\end{align}
where $\Delta$ is the anisotropy parameter. The interior metric
function $\lambda$ can be found from Eq.~\p{ve}:
\begin{align}
e^{-\lambda(r)}=1-\frac{2G}{r}m(r),\quad r<R, \label{lambda}
\end{align}
where $R$ is the radial coordinate at the surface of a star and
$m(r)$ is the mass enclosed in the sphere of radius $r$:
\begin{align}
m(r)=4\pi\int_0^r \varepsilon r^2 dr. \label{mr}
\end{align}
From Eq.~\p{pr}, with account of Eq.~\p{lambda}, one can find
\begin{align}
\nu\,'=2G\frac{m(r)+4\pi p_r r^3}{r(r-2Gm(r))}.\label{nupr}
\end{align}
Hence, the equation of hydrostatic equilibrium for a spherically
symmetric anisotropic star takes the form
\begin{align}
p_r^{\,\prime}=-G\frac{(\varepsilon+p_r)(m(r)+4\pi p_r
r^3)}{r(r-2Gm(r))}+\frac{2\Delta}{r}, \label{TOV}
\end{align}
which is the generalization of the Tolman–-Oppenheimer–-Volkoff
(TOV) equation with account of the local pressure anisotropy. In
order to solve Eq.~\p{TOV}, considered together with Eq.~\p{mr}, it
is necessary to set the EoS of the system,
$\varepsilon=\varepsilon(p_r)$, and to use a specific
model-dependent expression for $\Delta$, or to utilize some
approximation for it. Further we will assume that any physical
quantity or metric function in the interior of a star is free of
singularities. As follows from Eq.~\p{TOV}, the gradient
$p_r^{\,\prime}$ will be finite at $r=0$, if, at least,
$\Delta\propto r$ at $r\rightarrow0$. As the boundary condition to
Eq.~\p{TOV}, we will set the radial pressure at the center of a star
$p_r(0)=p_{r0}$, and will determine the radial coordinate $R$ at the
surface from the condition $p_r(R)=0$. The total mass then can be
determined as $M=m(R)$, assuming that $m(0)=0$. After finding the
radial pressure distribution $p_r(r)$ (together with the mass
distribution $m(r)$), the metric functions $\lambda(r),\nu(r)$ can
be determined from Eqs.~\p{lambda},\p{nupr}. At the boundary $r=R$,
the metric functions are matchable to the exterior vacuum
Schwarzschild metric:
\begin{align}
\lambda(R)=-\nu(R)=-\ln\biggl(1-\frac{2GM}{R}\biggr).\label{bcm}
\end{align}

\section{Generalized Lane--Emden equations}

Further, as the EoS of the system, we choose the polytropic EoS in
the form~\cite{ApJ65Tooper}:
\begin{align}
p_r=K\varrho^\gamma\equiv K\varrho^{1+\frac{1}{n}},\label{EoS}
\end{align}
where $\varrho$ is the mass (baryon) density, $K$ is some constant,
which can be, in principle, temperature dependent, $\gamma$ is the
polytropic exponent, $n$ is the polytropic index. Note that in some
works~\cite{ApJ64Tooper,Horedt04,APPB07Kinasiewicz,EPJC16Shojai} the
polytropic EoS is set in the form $p_r=K\varepsilon^\gamma$, which
will not be considered here. It is possible to show (see, e.g.,
Ref.~\cite{PRD09Read}) that for the
EoS~\p{EoS} 
the energy density $\varepsilon$ is related to the mass density
$\varrho$ and the radial pressure $p_r$ by the equation
\begin{align}
\varepsilon=\varrho+\frac{p_r}{\gamma-1}.\label{PEoS}
\end{align}

It is convenient to introduce the auxiliary dimensionless Lane-Emden
function $\theta$ according to the equations:
\begin{align}
p_r=p_{r0}\theta^{n+1},\quad \varrho=\varrho_0\theta^n,\label{theta}
\end{align}
where $\varrho_0$ is the central mass density. It follows from the
boundary conditions for the radial pressure $p_r$  that
\begin{align}
\theta(0)=1,\quad \theta(R)=0.\label{thetabc}
\end{align}
Then Eq.~\p{prpr} of hydrostatic equilibrium can be rewritten as
\begin{align}
2q_0(n+1)d\theta-\frac{4\Delta
dr}{\varrho_0r\theta^n}+\bigl(1+(n+1)q_0\theta\bigr)d\nu=0,\label{maine}
\end{align}
where $q_0\equiv\frac{p_{r0}}{\varrho_0}$. This equation can be
integrated to give
\begin{widetext}
\begin{equation*}
\nu=\nu_0-\ln\biggl(\frac{1+(n+1)q_0\theta}{1+(n+1)q_0}\biggr)^2+\frac{4}{\varrho_0}\int_0^r\frac{\Delta
dr}{r\theta^n\bigl(1+(n+1)q_0\theta\bigr)},
\end{equation*}
$\nu_0\equiv \nu(0)$  being the integration constant. In order to
find it, one can use the boundary condition~\p{bcm}. This gives
\begin{equation*}
\nu_0=\ln\frac{1-\frac{2GM}{R}}{\bigl(1+(n+1)q_0\bigr)^2}-\frac{4}{\varrho_0}\int_0^R\frac{\Delta
dr}{r\theta^n\bigl(1+(n+1)q_0\theta\bigr)}.
\end{equation*}
Therefore, the metric function $\nu(r)$ can be written as
\begin{align}
\nu(r)=\ln\frac{1-\frac{2GM}{R}}{\bigl(1+(n+1)q_0\theta\bigr)^2}-\frac{4}{\varrho_0}\int_r^R\frac{\Delta
dr}{r\theta^n\bigl(1+(n+1)q_0\theta\bigr)}.
\end{align}
Let us define the auxiliary function
\begin{align}
u(r)\equiv
\frac{m(r)}{M}=\frac{r}{2GM}\biggl(1-e^{-\lambda(r)}\biggr), \;
u(0)=0, u(R)=1, \label{u}
\end{align}
which, according to Eq.~\p{mr}, satisfies the differential equation
\begin{align}
Mu\,'=4\pi\varepsilon r^2.\label{upr}
\end{align}
Then, substituting $e^{-\lambda}$ from Eq.~\p{u}, and the derivative
$\nu\,'$ from Eq.~\p{maine} to Eq.~\p{pr}, and using Eq.~\p{upr},
one gets
\begin{gather}\label{preLE}
\frac{q_0(n+1)\theta\,'r}{1+(n+1)q_0\theta}\biggl(1-\frac{2GM}{r}u\biggr)+\frac{GMq_0\theta}{1+nq_0\theta}
u\,'\\
+\frac{GM}{r}u-\frac{2\Delta}{\varrho_0\theta^n\bigl(1+(n+1)q_0\theta\bigr)}\biggl(1-\frac{2GM}{r}u\biggr)=0.
\nonumber\end{gather}
Defining the dimensionless variable $\xi$ and dimensionless function
$\eta$ by the equations
\begin{align}
r=\alpha\xi, \quad\eta=\frac{M}{4\pi\varrho_0\alpha^3}u,\label{xi}
\end{align}
where $\alpha^2=\frac{q_0(n+1)}{4\pi G\varrho_0}$, Eqs.~\p{upr}  and
\p{preLE} can be rewritten as
\begin{align}
\frac{\xi-2(n+1)q_0\eta}{1+(n+1)q_0\theta}\biggl\{\xi\frac{d\theta}{d\xi}-\frac{2\Delta}{\varrho_0q_0(n+1)\theta^n
(1+(n+1)q_0\theta\bigr)}\biggr\}+\eta+q_0\xi^3\theta^{n+1}=0,\label{thetapr}\\
\frac{d\eta}{d\xi}=\xi^2\theta^n(1+nq_0\theta).\label{eta}
\end{align}
\end{widetext}
As follows from Eqs.~\p{thetabc} and \p{xi}, the boundary conditions
for the functions $\theta(\xi)$ and $\eta(\xi)$ read
\begin{align}
\theta(0)&=1,\; \theta(\xi_R)=0,\label{thetabounc}\\
\eta(0)&=0,\;
\eta(\xi_R)=\frac{M}{4\pi\varrho_0\alpha^3},\label{etabc}
\end{align}
where $\xi_R=R/\alpha$. Eqs.~\p{thetapr} and \p{eta} represent the
generalized  Lane-Emden equations for relativistic anisotropic
polytropes with the EoS~\p{EoS}, after solving which one can find
from Eqs.~\p{theta} and \p{xi} the radial distribution of the radial
pressure and mass in the interior of a spherically symmetric
relativistic anisotropic star. At $\Delta=0$, Eqs.~\p{thetapr} and
\p{eta} go over to the equations for the relativistic isotropic
polytropes~\cite{ApJ65Tooper}.

Note that for a given $q_0$, the radius and mass can be found as
functions of the constant $K$ in EoS~\p{EoS}:
\begin{align}
R=R^*q_0^\frac{1-n}{2}\xi_R,\;
M=M^*q_0^\frac{3-n}{2}\eta(\xi_R),\label{M*}
\end{align}
where the dependence on $K$ goes through the quantities $R^*$ and
$M^*$:
\begin{align}
R^*=\sqrt{\frac{n+1}{4\pi G}}K^\frac{n}{2}, \;
M^*=\frac{1}{\sqrt{4\pi}}\biggl(\frac{n+1}{G}\biggr)^\frac{3}{2}K^\frac{n}{2}.
\end{align}

The mass-radius relation has the form
\begin{align}
\frac{GM}{R}=(n+1)q_0\frac{\eta(\xi_R)}{\xi_R}. \label{MR}
\end{align}
\begin{widetext}
The total energy of a star is
\begin{align}
E=4\pi\int_0^R\varepsilon r^2dr=M^*q_0^\frac{3-n}{2}\eta(\xi_R).
\end{align}
The proper energy $E_0$ is the integral of the energy density over
the proper spatial volume~\cite{ApJ64Tooper}:
\begin{align}
E_0=4\pi\int_0^R \varepsilon
e^\frac{\lambda}{2}r^2dr=4\pi\varrho_0\alpha^3\int_0^{\xi_R}(1+nq_0\theta)\frac{\xi^2\theta^n}{\sqrt{1-
\frac{2q_0(n+1)}{\xi}\eta(\xi)}}d\xi.
\end{align}
The gravitational potential energy $\Omega$ is determined as
\begin{align}
\Omega=E-E_0=M^*q_0^\frac{3-n}{2}\eta(\xi_R)\biggl(1-\frac{1}{\eta(\xi_R)}
\int_0^{\xi_R}(1+nq_0\theta)\frac{\xi^2\theta^n}{\sqrt{1-
\frac{2q_0(n+1)}{\xi}\eta(\xi)}}d\xi\biggr).
\end{align}
The binding energy is the difference between the energy of the
particles scattered to infinity and the total energy of the
system~\cite{ApJ64Tooper}:
\begin{align}
E_B=E_{0g}-E, \;E_{0g}=4\pi\int_0^R\varrho e^\frac{\lambda}{2}r^2dr,
\end{align}
which reads
\begin{align}
E_B=M^*q_0^\frac{3-n}{2}\eta(\xi_R)\biggl(\frac{u_g(\xi_R)}{\eta(\xi_R)}-1\biggr),\;
u_g(\xi)=\int_0^{\xi}\frac{\xi^2\theta^n}{\sqrt{1-
\frac{2q_0(n+1)}{\xi}\eta(\xi)}}d\xi. \label{BE}
\end{align}
\end{widetext}

\section{Analytical solutions and numerical results for incompressible anisotropic fluid stars}

In order to solve the generalized Lane--Emden equations~\p{thetapr},
\p{eta} for the functions $\theta,\eta$ with the boundary
conditions~\p{thetabounc}, \p{etabc}, one needs to specify the
anisotropy parameter $\Delta$. This can be done within the given
model framework with the concrete  physical mechanism responsible
for the appearance of the local pressure anisotropy. The other
approach to study the effects of the pressure anisotropy is to set
the anisotropy parameter $\Delta$ in a phenomenological way, with
the use of some phenomenological ansatz for the anisotropy
parameter. This approach allows  to study general properties of
spherically symmetric relativistic anisotropic stars and  will be
followed in the given research. The conclusions obtained within this
approach are rather of a general character and independent of the
details of a specific physical mechanism.

In fact, we will follow the point of view, suggested in
Refs.~\cite{JHEP12Kim,EPJC15Shojai}, and which consists in setting
some additional differential relation between the unknown functions.
Namely, we will suppose that the presence of the anisotropy
parameter $\Delta$ doesn't change the general form of Lane-Emden
equations for relativistic isotropic stars~\cite{ApJ65Tooper}, but
can only change  the coefficients in these equations. Specifically,
we will assume that the parameter $\Delta$ and the metric function
$\nu$ are related by the differential equation
\begin{align}
-\frac{4\Delta
dr}{\varrho_0r\theta^n}+\bigl(1+(n+1)q_0\theta\bigr)d\nu=(1+\beta
q_0\theta)d\nu, \label{ansatz}
\end{align}
where $\beta$ is some real constant. Substituting Eq.~\p{ansatz}
into Eq.~\p{maine} and integrating it, one can obtain the metric
function $\nu(r)$ in the form
\begin{align}
\nu(r)=\ln\frac{1-\frac{2GM}{R}}{\bigl(1+\beta
q_0\theta\bigr)^\frac{2(n+1)}{\beta}}.\label{nuan}
\end{align}
Using Eqs.~\p{lambda} and \p{nuan}, and introducing the same
dimensionless variable $\xi$ and dimensionless function $\eta$ by
Eq.~\p{xi}, the modified Lane-Emden equations, corresponding to the
ansatz~\p{ansatz}, take the form
\begin{align}
\frac{\xi-2(n+1)q_0\eta}{1+\beta
q_0\theta}\xi\frac{d\theta}{d\xi}+\eta+q_0\xi^3\theta^{n+1}=0,\label{thetaprn}\\
\frac{d\eta}{d\xi}=\xi^2\theta^n(1+nq_0\theta)\label{etapr}
\end{align}
with the same boundary conditions~\p{thetabounc} and \p{etabc}. One
can see that the obtained Lane-Emden equations formally look as in
the isotropic case~\cite{ApJ65Tooper}, but with that difference that
the impact of the anisotropy parameter is reflected in the
coefficient $\beta$ (substituting the multiplier $(n+1)$).

In general case, the Lane-Emden equations~\p{thetaprn} and \p{etapr}
can be integrated only numerically, but the analytical solutions can
be found for incompressible anisotropic fluid stars, characterized
by the constant density $\varrho=\mathrm{const}$. At $n=0$, the
function $\eta(\xi)$, with account of the boundary condition
$\eta(0)=0$, is given by
\begin{align}
\eta(\xi)=\frac{\xi^3}{3}.\label{eta_xi}
\end{align}
The solution for the function $\theta(\xi)$ reads
\begin{align}
\frac{1+3q_0\theta}{1+\beta q_0\theta}=\pm\frac{1+3q_0}{1+\beta
q_0}\biggl(1-\frac{2q_0}{3}\xi^2\biggr)^\frac{3-\beta}{4}.
\end{align}
In the last equation, only the branch corresponding to the upper
plus sign, satisfies the boundary conditions~\p{thetabounc}, and the
respective solution is given by 
\begin{align}
\theta(\xi)=\frac{1}{q_0}\frac{(1+3q_0)\bigl(1-\frac{2q_0}{3}\xi^2\bigr)^\frac{3-\beta}{4}-(1+\beta
q_0)}{3(1+\beta
q_0)-\beta(1+3q_0)\bigl(1-\frac{2q_0}{3}\xi^2\bigr)^\frac{3-\beta}{4}}.
\end{align}
The positive root of the $\theta(\xi)$ reads
\begin{align}
\xi_R=\sqrt{\frac{3}{2q_0}\biggl[1-\biggl(\frac{1+\beta
q_0}{1+3q_0}\biggr)^{\frac{4}{3-\beta}}\biggr]}. \label{xi_R}
\end{align}

It is possible to check that the subradical function is nonnegative
at any $\beta$ satisfying the inequality $1+\beta q_0>0$.

Calculating the function $u_g(\xi)$ in Eq.~\p{BE} at $n=0$, the
binding energy of incompressible anisotropic fluid stars can be
written as
\begin{widetext}
\begin{align}
E_B=M^*\biggl\{-\frac{3}{4}\xi_R\sqrt{q_0\bigl(1-\frac{2q_0}{3}\xi_R^2\bigr)}+\frac{1}{2}\sqrt{\biggl(\frac{3}{2}\biggr)^3}
\arcsin\bigl(\sqrt{\frac{2q_0}{3}}\xi_R\bigr)-\frac{1}{3}\bigl(\sqrt{q_0}\xi_R\bigr)^3\biggr\}.\label{BEn0}
\end{align}
\end{widetext}
It can be verified that at $n=0$ the gravitational potential energy
is just  $\Omega=-E_B$.

\begin{figure*}[thb]
\begin{center}
\includegraphics[width=0.82\linewidth]{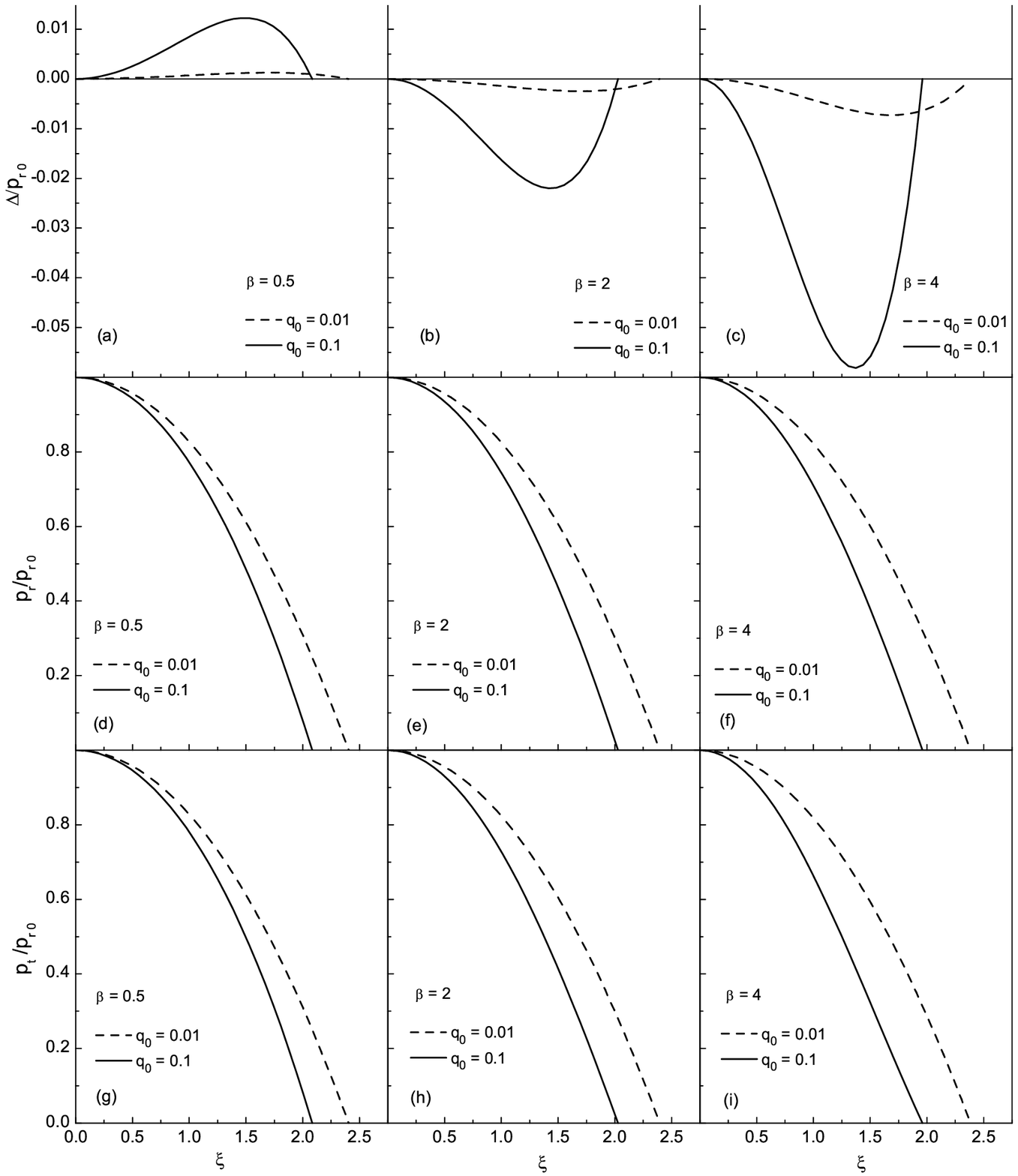}
\end{center}
\vspace{-2ex} \caption{The anisotropy parameter $\Delta$ (upper
row), the radial $p_r$ (middle row) and transverse $p_t$ (bottom
row) pressures at $n=0$, normalized to the central radial pressure
$p_{r0}$, as functions of the dimensionless variable $\xi$ for a set
of fixed values of $\beta$ and $q_0$.} \label{fig1}\vspace{-0ex}
\end{figure*}

Using the obtained analytical results for incompressible anisotropic
fluid stars, the basic quantities of interest can be represented
also in the graphical form. From Eq.~\p{ansatz}, one can find the
anisotropy parameter $\Delta$ at $n=0$ in terms of the dimensionless
variable $\xi$:
\begin{align}
\Delta(\xi)=\frac{p_{r0}\xi}{4}\bigl(1-\beta\bigr)\theta(\xi)\frac{\partial\nu}{\partial\xi}.
\label{delta_xi}\end{align} The radial pressure $p_r$, in turn, is
determined from Eq.~\p{theta}. Fig.~\ref{fig1} shows the ratios
$\Delta(\xi)/p_{r0}$,
 $p_r(\xi)/p_{r0}$ and $p_t(\xi)/p_{r0}$ at some fixed values of the
parameter $q_0=\frac{p_{r0}}{\varrho_0}$ and parameter $\beta$,
characterizing the impact of the pressure anisotropy. The important
peculiarity of our model ansatz~\p{ansatz} is that the anisotropy
parameter $\Delta$ vanishes both in the center and at the surface of
a star.  The anisotropy parameter $\Delta$ decreases with $\beta$
being positive at $\beta<1$ and being negative at $\beta>1$.

The radial pressure $p_{r}$ gradually decreases from its maximum
value $p_{r0}$ in the center till it vanishes at the surface of a
star. Also, $p_{r}$ decreases with increasing $\beta$. Qualitatively
the same behavior is demonstrated by the transverse pressure $p_t$.
The important feature of the model ansatz~\p{ansatz} is that not
only $p_r$, but also $p_t$ is positive in the interior of a star,
and vanishes at its surface. The positiveness of the radial $p_r$
and transverse $p_t$ pressures in the interior of a star guarantees
its mechanical stability. If some of these pressures becomes
negative, like the radial pressure $p_r$ in ultrastrong magnetic
fields, this leads to the appearance of the corresponding
instability; in the case of strong magnetic fields, this is the
longitudinal instability developed along the magnetic field
direction~\cite{Kh,FIKPS,PRD11Paulucci,IY_PRC11,IY_PLB12,IY_PAST12,JPG13IY}.
Vanishing of the transverse pressure $p_t$ at the surface of
anisotropic fluid stars is also important for their
stability~\cite{PRD11Paulucci,LNP13Ferrer,IJMPA14I,PRC15I}, although
in some studies this is not a required
feature~\cite{ApJ74Bowers,EPJC16Maurya,EPJC16Shojai}.

\begin{figure*}[thb]
\begin{center}
\includegraphics[width=0.82\linewidth]{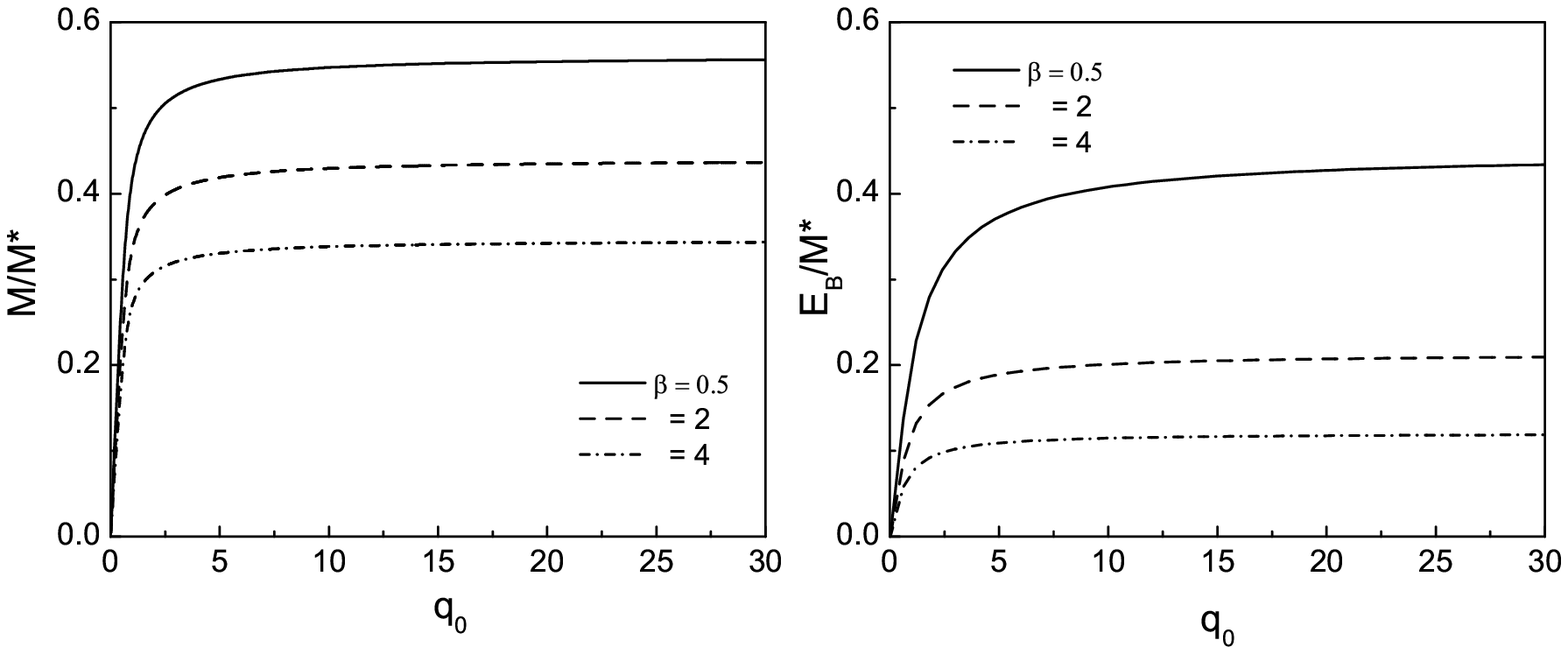}
\end{center}
\vspace{-2ex} \caption{The dimensionless mass $M/M^*$ and
dimensionless binding energy $E_B/M^*$ at $n=0$ as functions of the
parameter $q_0$ for the same set of fixed values of the parameter
$\beta$ as in Fig.~\ref{fig1}.} \label{fig2}\vspace{-0ex}
\end{figure*}

Fig.~\ref{fig2} shows the dimensionless mass $M/M^*$ and
dimensionless binding energy $E_B/M^*$ for incompressible
anisotropic fluid stars, determined according to Eqs.~\p{M*} and
\p{BEn0}, respectively, as functions of the parameter $q_0$ for the
set of fixed values of the parameter $\beta$. It is seen that both
quantities, first, rapidly increase with $q_0$ and then gradually
approach their asymptotic values, dependent on the given value of
$\beta$. The mass of a star decreases with increasing the parameter
$\beta$ at the given $q_0$, and, hence, the pressure anisotropy with
$\beta<1$ ($p_t>p_r$) leads to the increase of the mass of a star
compared to that in the isotropic case. The binding energy stays
always positive, as required by the stability of a star (although
not all stars with $E_B>0$ are stable, as will be shown in the next
section). The gravitational potential energy at $n=0$ differs only
by sign from the binding energy, $\Omega/M^*=-E_B/M^*$, and is
always negative.

\begin{figure}[thb]
\begin{center}
\includegraphics[width=8.6cm]{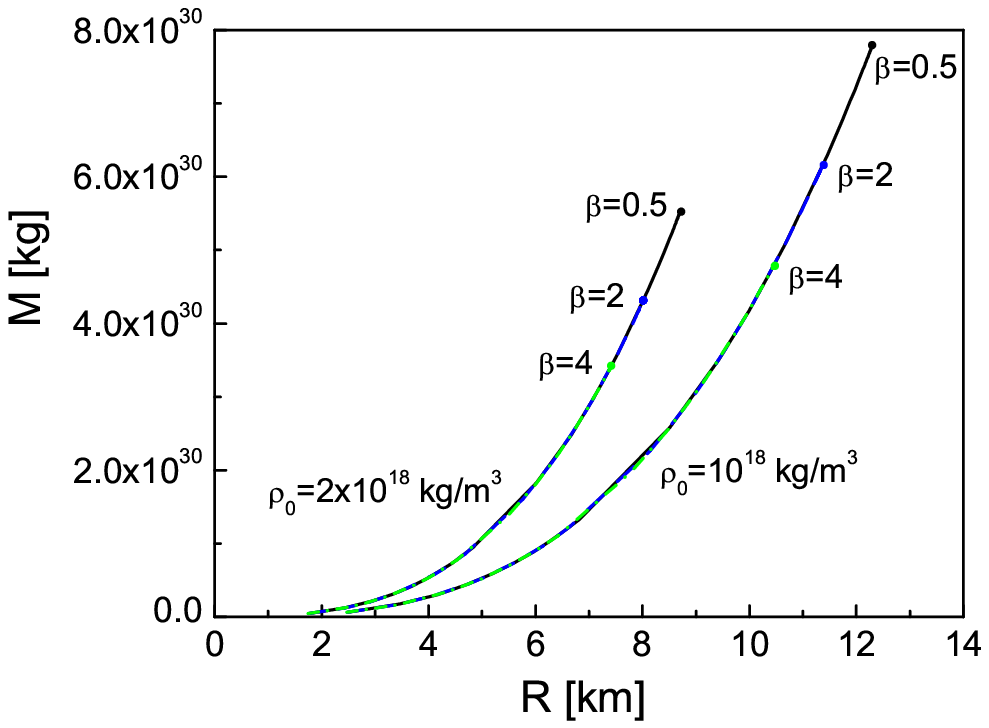}
\end{center}
\vspace{-2ex} \caption{(Color online) The mass--radius relation at
$n=0$ for two values of the central mass density,
$\varrho_0=10^{18}$~kg/m$^3$ and
$\varrho_0=2\times10^{18}$~kg/m$^3$, with the variable parameter
$q_0$ and fixed values of the parameter $\beta$: $\beta=0.5$ (black
curves), $\beta=2$ (blue curves) and $\beta=4$ (green curves). The
limiting masses for each configuration are shown by full dots.}
\label{fig3}\vspace{-0ex}
\end{figure}

Fig.~\ref{fig3} shows the mass--radius relation at $n=0$ for two
values of the central mass density, $\varrho_0=10^{18}$~kg/m$^3$ and
$\varrho_0=2\times10^{18}$~kg/m$^3$, obtained by varying the
parameter $q_0$ at fixed values of the parameter $\beta$. For the
specific central mass density, the masses and radii, corresponding
to different fixed  $q_0$ and $\beta$, are different. Nevertheless,
when $q_0$ varies at the given $\beta$, the current point moves
along almost the same curve for this specific central density
independently of $\beta$, the difference being only in the limiting
masses for  different $\beta$. Note that the maximum mass for
incompressible anisotropic fluid stars with the polytropic EoS 
cannot be reached at finite $q_0$, but only asymptotically at
$q_0\rightarrow\infty$. This is in contrast to the common behavior
when a mass-radius curve reaches a maximum value and then decreases
(cf. the mass-radius curves, for example, in
Refs.~\cite{HZS,JPG06Menezes} for strange quark stars). This result
does not rely on any numerical procedure, but just is the reflection
in the graphical form of the mass-radius relation~\p{MR} with the
analytically found  solutions of the modified Lane-Emden
equations~\p{thetaprn}, \p{etapr} at $n=0$, where the quantities
$\eta(\xi),\xi_R$ are given by Eqs.~\p{eta_xi}, \p{xi_R}.

The recent discovery of two heavy neutron stars PSR
J1614-2230~\cite{Nat10D} and PSR J0348+0432~\cite{Sci13A} with the
masses $M\sim 2M_{\odot}$ ($M_{\odot}\approx 1.989\times10^{30}$~kg
being the solar mass) set the corresponding constraint, to which any
reasonable EoS should satisfy. It is seen from Fig.~\ref{fig3} that,
despite its simplicity, the polytropic EoS for incompressible
anisotropic fluid stars can fulfill this constraint even without
reaching the maximum masses.

\section{Dynamical stability of incompressible anisotropic fluid stars}

Let us consider the stability of spherically symmetric anisotropic
stars with respect to  radial oscillations, assuming that they do
not violate the spherical symmetry. In the spherically symmetric
case, allowing for the motions in the radial direction, Einstein
equations~\p{Eeq} read \begin{widetext}
\begin{gather}
e^{-\lambda}\biggl(\frac{\lambda\,'}{r}-\frac{1}{r^2}\biggr)+\frac{1}{r^2}=8\pi
GT_0^0,\label{dve}\\
-e^{-\lambda}\biggl(\frac{\nu\,'}{r}+\frac{1}{r^2}\biggr)+\frac{1}{r^2}=8\pi
GT_1^1,\label{dpr}\\
-\frac{1}{2}e^{-\lambda}\biggl(\nu\,{''}-\frac{1}{2}\nu\,'\lambda\,'+\frac{1}{2}\nu\,'^2+
\frac{\nu\,'-\lambda\,'}{r}\biggr)+\frac{1}{2}e^{-\nu}\biggl(\ddot\lambda+
\dot\lambda(\dot\lambda-\dot\nu)\biggr)=8\pi GT_2^2=8\pi GT_3^3,
\label{dpt}\\
-\frac{e^{-\lambda}}{r}\dot\lambda=8\pi GT_0^1,\label{t01}
\end{gather}

where dot means the differentiation with respect to $t$. The radial
component of the equation $T_{i;k}^k=0$, expressing the vanishing of
the  divergence of the energy-momentum tensor, can be written as
\begin{align}
\dot T_1^0+
{T_1^1}^\prime+\frac{1}{2}T_1^0\bigl(\dot\nu+\dot\lambda\bigr)+\frac{\nu^{\,\prime}}{2}\bigl(T_1^1-T_0^0\bigr)+
\frac{2}{r}\bigl(T_1^1-T_2^2\bigr)=0.\label{ddiv}\end{align}
\end{widetext}
The energy--momentum tensor for a spherically symmetric anisotropic
star reads
\begin{align}
T_i^k=(\varepsilon+p_t)u_iu^k-p_t\delta_i^k+(p_r-p_t)s_is^k,\label{tik}
\end{align}
where $u^i=\frac{dx^i}{ds}$ is the fluid four-velocity, and $s^i$ is
the unit spacelike vector with the properties
\begin{align}
s^iu_i=0,\; s^is_i=-1.
\end{align}
For  the motions in the radial direction, the four-vectors $u^i$ and
$s^i$ have the structure
\begin{align}
u^i=\biggl(\frac{e^{-\frac{\nu}{2}}}{\sqrt{1-v^2e^{\lambda-\nu}}},\frac{ve^{-\frac{\nu}{2}}}
{\sqrt{1-v^2e^{\lambda-\nu}}},0,0\biggr),\label{ui}\\
s^i=\biggl(\frac{ve^{\frac{\lambda}{2}-\nu}}
{\sqrt{1-v^2e^{\lambda-\nu}}},\frac{e^{-\frac{\lambda}{2}}}{\sqrt{1-v^2e^{\lambda-\nu}}},0,0\biggr),
\end{align}
where $v=\frac{dr}{dt}$ is the velocity in the radial direction. In
the static limit, Eq.~\p{tik} for the energy-momentum tensor goes
over to Eq.~\p{emt}.

Further we will study the small radial oscillations and will
represent the unknown quantities as
\begin{gather}
\varepsilon=\varepsilon^0+\delta\varepsilon,\; p_r=p_{r}^0+\delta
p_r,\; p_t=p_{t}^0+\delta p_t,\label{pert}\\ \nu=\nu^0+\delta\nu,\;
\lambda=\lambda^0+\delta\lambda,
\end{gather}
where $\delta\varepsilon,\delta p_r,\delta
p_t,\delta\nu,\delta\lambda$ are small perturbations with respect to
the corresponding values at the state of hydrostatic equilibrium,
denoted by the upper index "0".  Also, for the small radial
oscillations we will consider that $v\ll1$. In the linear
approximation on the small perturbations, for the nonzero components
of the energy--momentum tensor we have
\begin{align}
T_0^0=\varepsilon,\;T_1^1=-p_r,\;T_2^2=T_3^3=-p_t,\label{emtp}\\T_0^1=(\varepsilon^0+p_{r}^0)v,
\; T_1^0=-e^{\lambda^0-\nu^0}T_0^1,\nonumber\end{align} where
quantities $\varepsilon,p_r$ and $p_t$ are given by Eq.~\p{pert}.
With account of Eq.~\p{emtp}, Eqs.~\p{dve} and \p{dpr} can be
written in the linearized form:
\begin{align}
\frac{\partial}{\partial
r}\bigl(re^{-\lambda^0}\delta\lambda\bigr)=8\pi
Gr^2\delta\varepsilon,\label{lve}\\
\frac{1}{r}e^{-\lambda^0}\bigl(\delta\nu\,'-\nu^{\,0\prime}\delta\lambda\bigr)=\frac{1}{r^2}e^{-\lambda^0}\delta\lambda+
8\pi G\delta p_r.\label{lpr}
\end{align}
The linearized form of Eq.~\p{t01} is
\begin{align}
-\frac{e^{-\lambda^0}}{r}\dot{\delta\lambda}=8\pi
G(\varepsilon^0+p_{r}^0)v.\label{lt01}
\end{align}
In the linear approximation, Eq.~\p{ddiv} reads \begin{widetext}
\begin{align}
-e^{\lambda^0-\nu^0}\bigl(\varepsilon^0+p_{r}^0\bigr)\dot v-\delta
p_r^{\,\prime}-\frac{1}{2}\bigl(\varepsilon^0+p_r^0\bigr)\delta\nu^\prime-
\frac{1}{2}\bigl(\delta\varepsilon+\delta
p_r\bigr)\nu^{0\prime}+\frac{2}{r}\bigl(-\delta p_r+\delta
p_t\bigr)=0.\label{lrdiv}
\end{align}
Following Ref.~\cite{ApJ64Chandrasekhar}, it is convenient to
introduce a "Lagrange displacement" $\psi$ by the equation
$v=\dot\psi$. Then integration of Eq.~\p{lt01} gives
\begin{align}
\frac{1}{r}e^{-\lambda_0}\delta\lambda=-8\pi
G(\varepsilon^0+p_{r}^0)\psi.\label{aux}
\end{align}
Note that in the state of hydrostatic equilibrium, as a consequence
of Eqs.~\p{ve} and \p{pr}, the following relationship holds true
\begin{align}
\frac{1}{r}e^{-\lambda_0}\bigl(\nu^{\,0\prime}+\lambda^{\,0\prime}\bigr)=8\pi
G(\varepsilon^0+p_{r}^0).\label{aux2}
\end{align}
With account of the last equation, Eq.~\p{aux} becomes
\begin{align}
\delta\lambda=-\psi\bigl(\nu^{\,0\prime}+\lambda^{\,0\prime}\bigr).\label{dlambda}
\end{align}
Taking into account  Eq.~\p{aux}, Eq.~\p{lve}  reads
\begin{align}
\delta\varepsilon=-\psi\varepsilon^{\,0\prime}-\psi
p_r^{\,0\prime}-\bigl(\varepsilon^0+p_{r}^0\bigr)\frac{1}{r^2}\frac{\partial}{\partial
r}\bigl(r^2\psi\bigr).\label{dvare}
\end{align}
In view of the condition of hydrostatic equilibrium~\p{prpr}, the
last equation can be written as 
\begin{align}
\delta\varepsilon=-\psi\varepsilon^{\,0\prime}-\bigl(\varepsilon^0+p_{r}^0\bigr)\frac{e^\frac{\nu^0}{2}}{r^2}\frac{\partial}{\partial
r}\bigl(r^2e^\frac{-\nu^0}{2}\psi\bigr)-\frac{2\psi}{r}\bigl(p_t^0-p_r^0\bigr).\label{dvef}
\end{align}
Next, considering Eq.~\p{lpr}, with account of Eqs.~\p{aux}
and~\p{aux2}, we get
\begin{align}
\bigl(\varepsilon^0+p_{r}^0\bigr)\delta\nu^{\,\prime}=\Bigl(\delta
p_r-\bigl(\varepsilon^0+p_{r}^0\bigr)\bigl(\nu^{\,0\prime}+\frac{1}{r}\bigr)\psi\Bigr)\bigl(\nu^{\,0\prime}+
\lambda^{\,0\prime}\bigr).\label{dnupr}
\end{align}
Now let us assume that all perturbations contain the
dependence on time only through the exponential factor $e^{i\omega
t}$, where $\omega$ is the frequency of radial oscillations. Then,
with account of Eq.~\p{dnupr}, Eq.~\p{lrdiv} can be written as
\begin{gather}
\omega^2
e^{\lambda^0-\nu^0}\bigl(\varepsilon^0+p_{r}^0\bigr)\psi=\delta
p_r^{\,\prime}+\Bigl(\nu^{\,0\prime}+\frac{\lambda^{\,0\prime}}{2}\Bigr)\delta
p_r\label{omega}\\
+\frac{\nu^{\,0\prime}}{2}\delta\varepsilon-\frac{1}{2}\bigl(\varepsilon^0+p_{r}^0\bigr)
\Bigl(\nu^{\,0\prime}+\frac{1}{r}\Bigr)\bigl(\nu^{\,0\prime}+
\lambda^{\,0\prime}\bigr)\psi+\frac{2}{r}\bigl(\delta p_r-\delta
p_t\bigl),\nonumber
\end{gather}
where all variations now stand for the amplitudes of the
corresponding quantities with the time dependence given by the
exponent $e^{i\omega t}$. Note that the variation
$\delta\varepsilon$ is expressed through the Lagrange displacement
$\psi$ by Eq.~\p{dvare}. In order to relate the perturbation of the
radial pressure $\delta p_r$ to $\psi$, we will assume, following
Ref.~\cite{ApJ64Chandrasekhar}, the conservation of the total baryon
number. The corresponding continuity equation in general relativity
reads
\begin{align}
\frac{\partial}{\partial x^k}\bigl(n_bu^k\bigr)+n_bu^k\frac{\partial
\ln\sqrt{-g}}{\partial x^k}=0,
\end{align}
where $n_b$ is the baryon number density, $g$ is the determinant of
the metric tensor,
$g\equiv\det||g_{ik}||=-\exp(\nu+\lambda)r^4\sin^2\theta$. Taking
into account Eq.~\p{ui}, the last equation in the linear
approximation on the small perturbations takes the form
\begin{align}
e^{-\frac{\nu^0}{2}}\dot{\delta
n_b}+\frac{1}{r^2}\frac{\partial}{\partial
r}\bigl(n_b^0e^{-\frac{\nu^0}{2}}vr^2\bigr)+\frac{1}{2}n_b^0e^{-\frac{\nu^0}{2}}\dot{\delta\lambda}+
\frac{1}{2}n_b^0e^{-\frac{\nu^0}{2}}v\bigl(\nu^{\,0\prime}+\lambda^{\,0\prime}\bigr)=0,
\end{align}

where $n_b^0(r)$ is the baryon number density in the state of
hydrostatic equilibrium. Since $v=\dot\psi$, this equation can be
integrated and one gets
\begin{align}
\delta n_b+\frac{e^{\frac{\nu^0}{2}}}{r^2}\frac{\partial}{\partial
r}\bigl(n_b^0e^{-\frac{\nu^0}{2}}r^2\psi\bigr)+
\frac{1}{2}n_b^0\bigl(\delta\lambda+\psi\bigl(\nu^{\,0\prime}+\lambda^{\,0\prime}\bigr)\bigr)=0.\label{dn}
\end{align}
In view of Eq.~\p{dlambda}, the last term in Eq.~\p{dn} vanishes and
one obtains
\begin{align}
\delta n_b=-\frac{e^{\frac{\nu^0}{2}}}{r^2}\frac{\partial}{\partial
r}\bigl(n_b^0e^{-\frac{\nu^0}{2}}r^2\psi\bigr).\label{dnf}
\end{align}
Let the EoS of the system have the following general structure:
$n_b=n_b(\varepsilon,p_r)$. Then it follows from Eqs.~\p{dvef} and
\p{dnf} that in the linear approximation
\begin{align}
\delta p_r=-p_r^{0\prime}\psi-\gamma
p_r^0\frac{e^{\frac{\nu^0}{2}}}{r^2}\frac{\partial}{\partial
r}\bigl(e^{-\frac{\nu^0}{2}}r^2\psi\bigr)-\frac{2}{r}\frac{\partial
p_r}{\partial\varepsilon}\bigl(p_t^0-p_r^0\bigr)\psi,\label{dprf}
\end{align}
where the adiabatic coefficient $\gamma$ is determined by
\begin{align}
\gamma=\frac{1}{p_r\frac{\partial n_b}{\partial
p_r}}\Bigl(n_b-(\varepsilon+p_r)\frac{\partial
n_b}{\partial\varepsilon}\Bigr),
\end{align}
and, analogously to Ref.~\cite{ApJ64Chandrasekhar}, is considered to
be  a constant for the matter inside a star. For the polytropic
EoS~\p{PEoS}, $\frac{\partial p_r}{\partial\varepsilon}=\gamma-1$.
Substituting expressions~\p{dvef} and \p{dprf} for
$\delta\varepsilon$ and $\delta p_r$ to Eq.~\p{omega}, and using the
field equation~\p{pt} and equation~\p{prpr} of hydrostatic
equilibrium, in the case of the polytropic EoS one gets
\begin{gather}
\omega^2
e^{\lambda^0-\nu^0}\bigl(\varepsilon^0+p_{r}^0\bigr)\psi=\frac{2\psi}{r}p_r^{\,0\prime}
-
\frac{2\psi}{r}\Bigl(\gamma\bigl(\nu^{\,0\prime}+\frac{\lambda^{\,0\prime}}{2}+\frac{2}{r}\bigr)+
\frac{2}{r}\Bigr)\bigl(p_t^0-p_r^0\bigr) \label{eigfr}\\+8\pi
Ge^{\lambda^0}p_t^0\bigl(\varepsilon^0+p_r^0\bigr)\psi -
\gamma\frac{d}{dr}\Bigl(\frac{2}{r}\bigl(p_t^0-p_r^0\bigr)\psi\Bigr) 
-\frac{\psi}{\varepsilon^0+p_r^0}\Bigl(p_r^{\,0\prime}-
\frac{2}{r}\bigl(p_t^0-p_r^0\bigr)\Bigr)^2 \nonumber\\ - \gamma
e^{-(\nu^0+\frac{\lambda^0}{2})}\frac{d}{dr}\Bigl(e^\frac{3\nu^0+
\lambda^0}{2}\frac{p_r^0}{r^2}\frac{d}{dr}\bigl(r^2e^{-\frac{\nu^0}{2}}\psi\bigr)\Bigr)
-\frac{2}{r}\Bigl(\gamma
p_r^0\frac{e^\frac{\nu^0}{2}}{r^2}\frac{d}{dr}\bigl(r^2e^{-\frac{\nu^0}{2}}\psi\bigr)+\delta
p_t\Bigr).\nonumber
\end{gather}

Solutions of Eq.~\p{eigfr} for the frequencies of radial
oscillations should be sought under the boundary conditions
\begin{align}
\psi(r=0)=0,\; \delta p_r(r=R)=0. \label{bc}
\end{align}
In order to get the variational basis for finding the frequencies
$\omega$, let us multiply both parts of Eq.~\p{eigfr} on
$r^2\psi\exp\bigl(\frac{\nu^0+\lambda^0}{2}\bigr)$ and integrate
over the range of $r$. We will write the corresponding equation
already for incompressible fluid stars  ($n=0$), when the polytropic
exponent $\gamma\rightarrow\infty$. Omitting the upper indices zero
as no longer necessary, one gets
\begin{gather}
\omega^2\int_0^R
e^{\frac{3\lambda-\nu}{2}}\bigl(\varepsilon+p_{r}\bigr)r^2\psi^2\,dr=
\gamma\int_0^Re^{\frac{\lambda+3\nu}{2}}
\frac{p_r}{r^2}\Bigl(\frac{d}{dr}\bigl(r^2e^{-\frac{\nu}{2}}\psi\bigr)\Bigr)^2\,dr
\label{vareq}\\-
\gamma\int_0^Re^{\frac{\lambda+\nu}{2}}r^2\psi\frac{d}{dr}\Bigl(\frac{2}{r}\bigl(p_t-p_r\bigr)\psi\Bigr)\,dr
 -2\gamma\int_0^Re^{\frac{\lambda+\nu}{2}}r\psi^2\Bigl(
\nu^{\,\prime}+\frac{\lambda^{\,\prime}}{2}+\frac{2}{r}
\Bigr)\bigl(p_t-p_r\bigr)\nonumber\\
 -2\gamma\int_0^Re^{\frac{\lambda}{2}+\nu}\psi\frac{
 p_r}{r}\frac{d}{dr}\bigl(r^2e^{-\frac{\nu}{2}}\psi\bigr)\,dr.
\nonumber
\end{gather}
In the variational equation~\p{vareq}, the Lagrange displacement
$\psi$ should be chosen such that $\omega^2$ is minimized. If all
frequencies of  radial oscillations are real, a spherically
symmetric anisotropic star is dynamically stable; if some frequency
appears to be imaginary, the configuration is unstable. A sufficient
condition for the occurrence of the dynamical instability  is
vanishing of the right--hand side of Eq.~\p{vareq} for some trial
form of the Lagrange displacement $\psi$ satisfying the boundary
conditions.

Let us introduce, following Ref.~\cite{ApJ65Tooper}, the auxiliary
function $\chi=e^{-\frac{\nu}{2}}\psi$. After changing the
integration variable in Eq.~\p{vareq} according to Eq.~\p{xi},
substituting $p_r=q_0\varrho_0\theta$, $\varepsilon=\varrho_0$, and
using Eq.~\p{delta_xi} for the anisotropy parameter $\Delta=p_t-p_r$
and expressions for the metric functions at $n=0$:
\begin{align}
e^{-\lambda}=1-\frac{2q_0\eta(\xi)}{\xi}=1-\frac{2q_0\xi^2}{3},\;
e^\nu=\frac{1-\frac{2GM}{R}}{\bigl(1+\beta
q_0\theta\bigr)^\frac{2}{\beta}},
\end{align}
Eq.~\p{vareq}  takes the form
\begin{gather}
\frac{\omega^2}{\omega_0^2}\frac{1}{1-\frac{2GM}{R}}
\int_0^{\xi_R}\frac{(1+q_0\theta)\xi^2\chi^2}{\bigl(1-\frac{2q_0\xi^2}{3}\bigr)^\frac{3}{2}}\frac{d\xi}
{(1+\beta
q_0\theta)^\frac{1}{\beta}}=\gamma\int_0^{\xi_R}\frac{\theta\Bigl(\frac{d}{d\xi}\bigl(\xi^2\chi\bigr)\Bigr)^2}{\xi^2\bigl(1-\frac{2q_0\xi^2}{3}\bigr)^\frac{1}{2}}
\frac{d\xi}{(1+\beta
q_0\theta)^\frac{3}{\beta}}\label{dyneq}\\-\frac{\gamma(1-\beta)}{2}\int_0^{\xi_R}\frac{\xi^2\chi}
{\bigl(1-\frac{2q_0\xi^2}{3}\bigr)^\frac{1}{2}}\frac{d}{d\xi}\Bigl(\frac{\nu'\chi\theta}
{(1+\beta q_0\theta)^\frac{1}{\beta}}\Bigr)\frac{d\xi} {(1+\beta
q_0\theta)^\frac{2}{\beta}} \nonumber
\\-\frac{\gamma(1-\beta)}{2}\int_0^{\xi_R}\frac{\xi^2\chi^2\nu'\theta\bigl(\nu'+\frac{\lambda'}{2}+\frac{2}{\xi}\bigr)}
{\bigl(1-\frac{2q_0\xi^2}{3}\bigr)^\frac{1}{2}}\frac{d\xi} {(1+\beta
q_0\theta)^\frac{3}{\beta}}-2\gamma\int_0^{\xi_R}
\frac{\chi\theta\frac{d}{d\xi}\bigl(\xi^2\chi\bigr)}
{\xi\bigl(1-\frac{2q_0\xi^2}{3}\bigr)^\frac{1}{2}}\frac{d\xi}
{(1+\beta q_0\theta)^\frac{3}{\beta}},
 \nonumber
\end{gather}
\end{widetext}
where $\omega_0^2=4\pi\varrho_0G$. Let us use the trial functions of
the form
\begin{align}
\chi_1=e^{-\frac{\nu}{2}}\xi,\quad \chi_2=\sqrt{\xi}.
\end{align}
Then for each given $\beta$ we will try to find such $q_{0c}$ at
which the right-hand side of Eq.~\p{dyneq} vanishes, and, hence, the
dynamical instability for an incompressible anisotropic fluid star
occurs at $q_0>q_{0c}$.

\begin{table}[tbh]
\caption{The critical values of  the parameter $q_{0}$ for the
appearance of the dynamical instability of an incompressible
anisotropic fluid star at different values of the parameter $\beta$
and two types of the trial functions used in the calculations.}
\begin{center}
\begin{tabular}{c|cc}\hline
\multicolumn{1}{c|}{$\beta$}& \multicolumn{2}{c}{$q_{0c}$ evaluated
with the trial function} \\ \cline{2-3} &
\multicolumn{1}{l|}{\hspace{0.8cm}$\chi_1=e^{-\frac{\nu}{2}}\xi$\hspace{0.8cm}
} &
\multicolumn{1}{c}{$\chi_2=\sqrt{\xi}$} \\ \hline 0.1 & \multicolumn{1}{c|}{1.391} & - \\
{0.3} & \multicolumn{1}{c|}{1.796} & - \\
{0.5} & \multicolumn{1}{c|}{2.526} & - \\
0.7 & \multicolumn{1}{c|}{4.210} & - \\
0.9 & \multicolumn{1}{c|}{11.646} & - \\
\hline
\end{tabular}
\end{center}
 \label{table1}
\end{table}

The results of calculations are presented in Table~\ref{table1}. The
most important conclusion is that there are solutions for $q_{0c}$
in the case of the trial function $\chi_1$ at $\beta<1$, i.e., for
$\Delta=p_t-p_r>0$ (and there are no solutions at $\beta>1$). This
means that the local pressure anisotropy with $p_t>p_r$ can affect
the dynamical stability of spherically symmetric incompressible
fluid stars, the result which is in contrast to the conclusion for
incompressible isotropic fluid stars with the polytropic EoS~\p{EoS}
in Ref.~\cite{ApJ65Tooper}, which are stable against radial
oscillations.

\begin{figure}[thb]
\begin{center}
\includegraphics[width=8.6cm]{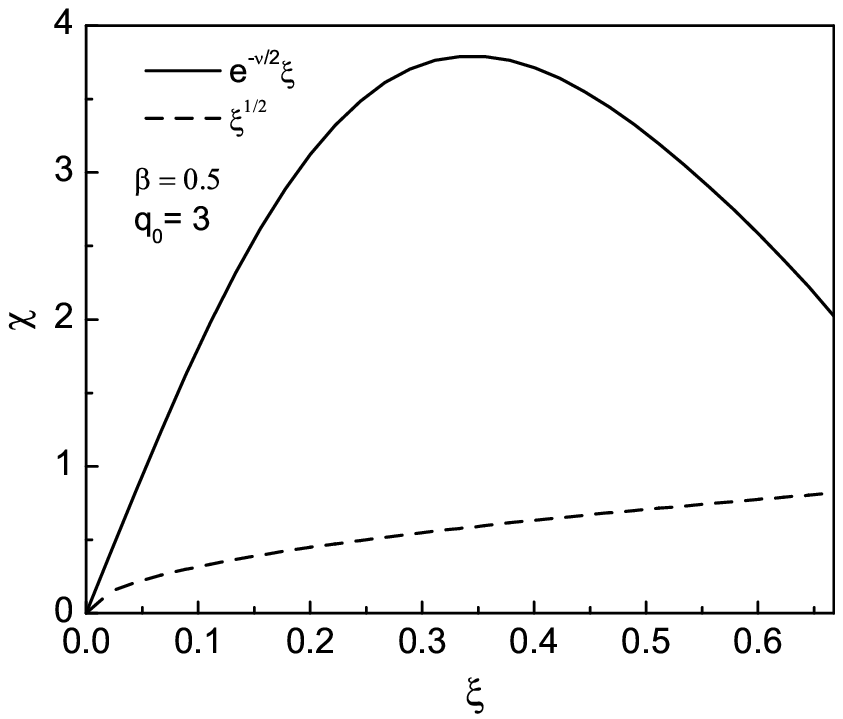}
\end{center}
\vspace{-2ex} \caption{The trial functions
$\chi_1=e^{-\frac{\nu}{2}}\xi$ and $\chi_2=\sqrt{\xi}$ at
$\beta=0.5, q_0=3$.} \label{fig4}\vspace{-0ex}
\end{figure}

It is seen also that the choice of the trial function does matter:
the use of $\chi_1$ allows us to find the critical value $q_{0c}$
for  $\beta<1$ while the right-hand side of Eq.~\p{dyneq} does not
vanish for the trial function $\chi_2$ at any $\beta$.
Fig.~\ref{fig4} shows the behavior of these trial functions at
$\beta=0.5, q_0=3$. While the derivative $\chi_2'(\xi)$ is always
positive, the derivative $\chi_1'(\xi)$ changes sign in the interval
$[0,\xi_R]$, and, hence, the subintegral functions containing
$\chi'(\xi)$ contribute qualitatively differently  to the respective
integrals for $\chi_1(\xi)$ and $\chi_2(\xi)$.

If dynamical instability occurs at $\beta<1$, the question naturally
arises: is the mass of an incompressible anisotropic fluid star
still compatible with the two--solar--mass constraint at the moment
of the appearance of dynamical instability at the critical value
$q_{0c}$? Table~\ref{table2} shows the values of the mass of an
incompressible anisotropic fluid star at the critical value $q_{0c}$
in the case of the trial function $\chi_1(\xi)$. It is seen that for
both values of the central mass density $\varrho_0$, used
in the calculations, 
the two--solar--mass constraint is still satisfied at the moment of
the appearance of dynamical instability.

\begin{widetext}
\begin{center}
\begin{table}[thb]
\caption{The mass $M_c$ of an incompressible anisotropic fluid star
at the critical value $q_{0c}$ for the appearance of dynamical
instability in the case of the trial function $\chi_1(\xi)$.}
\begin{center}
\begin{tabular}{c|ccccc}\hline
\multicolumn{1}{c|}{$\varrho_0$, kg/m$^3$}&
\multicolumn{5}{c}{\textrm{$M_{c}$ evaluated at the critical value
$q_{0c}$, kg }} \\
\cline{2-6} & \multicolumn{1}{l|}{$\beta=0.1$ } &
\multicolumn{1}{c|}{$\beta=0.3$} & \multicolumn{1}{c|}{$\beta=0.5$}
& \multicolumn{1}{c|}{$\beta=0.7$} & \multicolumn{1}{c}{$\beta=0.9$}
\\ \hline 10$^{18}$ & \multicolumn{1}{c|}{6.999$\times10^{30}$} & \multicolumn{1}{c|}{7.030$\times10^{30}$} &
\multicolumn{1}{c|}{7.064$\times10^{30}$} &
\multicolumn{1}{c|}{7.097$\times10^{30}$} &
\multicolumn{1}{c}{7.122$\times10^{30}$}
\\
$2\times 10^{18}$ & \multicolumn{1}{c|}{4.949$\times10^{30}$} &
\multicolumn{1}{c|}{4.971$\times10^{30}$} &
\multicolumn{1}{c|}{4.995$\times10^{30}$}
&\multicolumn{1}{c|}{5.018$\times10^{30}$} &
\multicolumn{1}{c}{5.036$\times10^{30}$}
\\
\hline
\end{tabular}
\end{center}
 \label{table2}
\end{table}
\end{center}
\end{widetext}

In summary, we have considered spherically symmetric relativistic
stars with the polytropic equation of state, which possess the local
pressure anisotropy, within  the framework of general relativity.
The generalized Lane-Emden equations have been derived for the case
of the arbitrary anisotropy parameter $\Delta=p_t-p_r$. In this
research, in order to study the effects of the pressure anisotropy,
we follow a phenomenological approach, in which the anisotropy
parameter is set with the help of some phenomenological ansatz.  The
conclusions obtained within this approach are rather of a general
character and independent of the details of a specific physical
mechanism. Specifically, the generalized Lane-Emden equations are
applied to the special ansatz~\p{ansatz} for the anisotropy
parameter $\Delta$ in the form of the differential relation between
$\Delta$  and the metric function $\nu$.
 The
analytical solutions of the obtained equations have been  found for
incompressible fluid stars and then used for getting their
mass-radius relation, gravitational and binding energy. It has been
clarified that the pressure anisotropy with $p_t>p_r$ leads to the
increase of the mass of a star compared to that in the isotropic
case, and this factor can be helpful in explaining the observational
data of heavy compact stars with the mass $M\sim 2M_{\odot}$. Also,
following the Chandrasekhar variational
approach~\cite{ApJ64Chandrasekhar}, the dynamical stability of
 incompressible anisotropic fluid stars with
the polytropic EoS against radial oscillations has been studied. It
has been shown that the local pressure anisotropy with $p_t>p_r$ can
make the incompressible fluid stars unstable with respect to radial
oscillations, in contrast to incompressible isotropic fluid stars
with the polytropic EoS which are dynamically
stable~\cite{ApJ65Tooper}. As shown in the model calculations of the
given work, if dynamical instability occurs, the mass of an
incompressible anisotropic fluid star at the moment of the
appearance of instability is still compatible with the
two--solar--mass constraint.

Note that in the interior of magnetars -- strongly magnetized
neutron or quark stars, magnetic fields of about $10^{18}$~G, or
even larger, can potentially occur. Such strong magnetic fields can
produce the substabtial pressure anisotropy with
$p_t>p_r$~\cite{Kh,FIKPS,PRD11Paulucci,LNP13Ferrer,IY_PRC11,IY_PLB12,IY_PAST12,JPG13IY,IJMPA14I,PRC15I,JPCS15I,PRD14Chu,PRD15Chu},
and, hence, can cause the dynamical instability of a magnetized
stellar object.


\section*{References}
\begin{thebibliography}{99}


\bibitem{ApJ74Bowers} R. L. Bowers and E. P. T. Liang, \Journal{\apj}{188}{657}{1974}.
\bibitem{CQG11Horvat}  D. Horvat, S. Ilijic and A. Marunovic, \Journal{\cqg}{28}{025009}{2011}.
\bibitem{PRD11Harko} T. Harko and F. S. N. Lobo, \Journal{\prd}{83}{124051}{2011}.
\bibitem{PRD13'87Herrera} L. Herrera, and W. Barreto, \Journal{\prd}{87}{087303}{2013}.
\bibitem{EPJC15Shojai} F. Shojai, M.R. Fazel, A. Stepanian, and M. Kohandel,
\Journal{\epjc}{75}{250}{2015}.
\bibitem{ARAA72Ruderman}M. Ruderman, Annu. Rev. Astron.
Astrophys. {\bf 10}, 427 (1972).
\bibitem{PR97Herrera} L. Herrera, and N. O. Santos, \Journal{\pr}{286}{53}{1997}.
\bibitem{PRD04Herrera} L. Herrera, A. Di Prisco, J. Martin, J. Ospino, N.O. Santos, and O.
Troconis, \Journal{\prd}{69}{084026}{2004}.
\bibitem{CQG05Cattoen} C. Cattoen, T. Faber, M. Visser, \Journal{\cqg}{22}{4189}{2005}.
\bibitem{PRD13Herrera} L. Herrera, and W. Barreto, Phys. Rev. D {\bf 88}, 084022 (2013).
\bibitem{EPJC16Maurya} S. K. Maurya, Y. K. Gupta, B. Dayanandan, and S.
Ray, \Journal{\epjc}{76}{266}{2016}.
\bibitem{EPJC16Shojai} F. Shojai,  A. Stepanian, and M. Kohandel,
\Journal{\epjc}{76}{347}{2016}.
\bibitem{PRL02Muther} H. M\"uther and A. Sedrakian, \Journal{\prl}{88}{252503}{2002}.
\bibitem{PRC03Muther} H. M\"uther and A. Sedrakian, \Journal{\prc}{67}{015802}{2003}.
\bibitem{PTP93Takatsuka}
T. Takatsuka and R. Tamagaki, \Journal{\ptps}{112}{27}{1993}.
 \bibitem{PRC98Baldo}
M. Baldo, O. Elgaroy, L. Engvik, M. Hjorth-Jensen and H.-J. Schulze,
\Journal{\prc}{58}{1921}{1998}.
\bibitem{PRC02IR} A.A.  Isayev and G. R\"opke, \Journal{\prc}{66}{034315}{2002}.
\bibitem{NPA03Zverev} M. V. Zverev, J. W. Clark  and V. A. Khodel, \Journal{\npa}{720}{20}{2003}.
\bibitem{EPL08Zuo} W. Zuo, A. J. Mi, C. X. Cui, and U. Lombardo , \Journal{\epl}{84}{32001}{2008}.

\bibitem{PRC02I} A.A.  Isayev, \Journal{\prc}{65}{031302}{2002}.
\bibitem{RMP04Casalbuoni} R. Casalbuoni and G. Nardulli, \Journal{\rmp}{76}{263}{2004}.


\bibitem{Kh} V. R. Khalilov, \Journal{\prd}{65}{056001}{2002}. 


\bibitem{FIKPS} E. J. Ferrer, V. de la Incera, J. P. Keith et al., \Journal{\prc}{82}{065802}{2010}.
\bibitem{PRD11Paulucci}
L. Paulucci, E. J. Ferrer,  V. de la Incera, and J. E. Horvath,
\Journal{\prd}{83}{043009}{2011}.
\bibitem{LNP13Ferrer}E. J. Ferrer, and V. de la Incera, \Journal{\lnp}{871}{399}{2013}.
\bibitem{IY_PRC11}  A.A. Isayev, and J. Yang, \Journal{\prc}{84}{065802}{2011}. 
\bibitem{IY_PLB12}  A.A. Isayev, and J. Yang, \Journal{\plb}{707}{163}{2012}. 
\bibitem{IY_PAST12} A. A. Isayev and J. Yang, \Journal{\past}{1}{11}{2012}.
\bibitem{JPG13IY} A.A.  Isayev, and  J. Yang, \Journal{\jpg}{40}{035105}{2013}. 
\bibitem{IJMPA14I} A.A.  Isayev, \Journal{\ijmpa}{29}{1450173}{2014}.
\bibitem{PRC15I} A.A.  Isayev, \Journal{\prc}{91}{015208}{2015}.
\bibitem{JPCS15I} A.A.  Isayev, \Journal{\jpcs}{607}{012013}{2015}.
\bibitem{PRD14Chu} P.C. Chu, L.W. Chen, and X. Wang, \Journal{\prd}{90}{063013}{2014}.
\bibitem{PRD15Chu}
P. C. Chu, X. Wang, L. W. Chen, M. Huang, 
\Journal{\prd}{91}{023003}{2015}.

\bibitem{PRD10Varela} V. Varela, F. Rahaman, S. Ray, K. Chakraborty, and M. Kalam, \Journal{\prd}{82}{044052}{2010}.

\bibitem{ApJ64Tooper} R.F. Tooper, \Journal{\apj}{140}{434}{1964}.
\bibitem{ApJ65Tooper} R.F. Tooper, \Journal{\apj}{142}{1541}{1965}.

\bibitem{Horedt04} G.P. Horedt, {\it Polytropes - Applications in Astrophysics and
Related Fields}, Astrophysics and Space Science Library, Vol. 306
(Kluwer Academic, Dordrecht, 2004).


\bibitem{JMP98Nilsson} U.S. Nilsson, C. Uggla, and M. Marklund, \Journal{\jmp}{39}{3336}{1998}.

\bibitem{CQG03Heinzle} J.M. Heinzle, N. R\"ohr, and C. Uggla, \Journal{\cqg}{20}{4567}{2003}.

\bibitem{APPB07Kinasiewicz} B. Kinasiewicz, and P. Mach, \Journal{\appb}{38}{39}{2007}.

\bibitem{PRD09Read} J. S. Read, B. D. Lackey, B. J. Owen, and J. L. Friedman,
\Journal{\prd}{79}{124032}{2009}.

\bibitem{Pram12Thirukkanesh} S. Thirukkanesh, F.C. Ragel, \Journal{\pram}{78}{687}{2012}.

\bibitem{GRG89Maharaj} S.D. Maharaj, and R.Maartens, \Journal{\grg}{21}{899}{1989}.

\bibitem{GRG94Gokhroo} M.K. Gokhroo, and A.L. Mehra, \Journal{\grg}{26}{75}{1994}.

\bibitem{MNRAS07Sharma} R. Sharma, and S.D. Maharaj, \Journal{\mnras}{375}{1265}{2007}.

\bibitem{ApJ64Chandrasekhar} S. Chandrasekhar, \Journal{\apj}{140}{417}{1964}.

\bibitem{JHEP12Kim} K. K. Kim, Y. Kim, and I. J. Shin,
\Journal{\jhep}{11}{045}{2012}.

\bibitem{HZS} P. Haensel, J. Zdunik,  R. Schaeffer, \Journal{\aap}{160}{121}{1986}.

\bibitem{JPG06Menezes}D. P. Menezes, C. Providencia, and D. B. Melrose, \Journal{\jpg}{32}{1081}{2006}.

\bibitem{Nat10D} P. Demorest, T. Pennucci, S. M. Ransom et al., \Journal{\nat}{467}{1081}{2010}.

\bibitem{Sci13A} J. Antoniadis, P. C. C. Freire, N. Wex et al., \Journal{\sci}{340}{6131}{2013}.

\end{thebibliography}
\end{document}